# Cocoa pollination, biodiversity-friendly production, and the global market

Thomas Cherico Wanger[1,2,3,4*], Francis Dennig[5,6], Manuel Toledo-Hernández[1,2,3], Teja Tscharntke[2], Eric F. Lambin[7,8]

[1]Sustainable Agricultural Systems & Engineering, School of Engineering, Westlake University, China. [2]Agroecology, University of Göttingen, Germany. [3]GlobalAgroforestryNetwork.org, China. [4]Key Laboratory of Coastal Environment and Resources of Zhejiang Province, Westlake University, China. [5]CASBS, Stanford University, USA. [6]Yale-NUS College, Singapore. [7]School of Earth, Energy & Environment Sciences and Woods Institute for the Environment, Stanford University, USA. [8]Georges Lemaître Earth and Climate Research Centre, Earth and Life Institute, Université catholique de Louvain, Belgium.

*Thomas Cherico Wanger

**Email:** tomcwanger@gmail.com

**Abstract**

Production of cocoa, the third largest trade commodity globally has experienced climate-related yield stagnation since 2016 (ICCO 2020), forcing farmers to expand production in forested habitats and to shift from nature-friendly agroforestry systems to intensive monocultures. The goal for future large-scale cocoa production combines high yields with biodiversity-friendly management into a climate-adapted (Blaser et al. 2018) 'smart agroforestry system' (SAS). As pollination limitation is a key driver of global production, we use data of more than 150,000 cocoa farms and results of hand pollination experiments to show that manually enhancing cocoa pollination ('manual pollination') can produce SAS. Manual pollination can triple farm yields and double farmers' annual profit in the major producer countries Ivory Coast, Ghana, and Indonesia, and can increase global cocoa supplies by up to 13%. We propose a win-win scenario to mitigate negative long-term price and socioeconomic effects, whereby manual pollination compensates only for yield losses resulting from climate and disease-related decreases in production area and conversion of monocultures into agroforestry systems. Our results highlight that yields in biodiversity–friendly and climate-adapted SAS can be similar to yields currently only achieved in monocultures. Adoption of manual pollination could be achieved through wider implementation of eco-certification standards, carbon markets, and zero deforestation pledges (Lambin et al. 2018).

**Significance Statement**
Climate change and disease outbreaks are causing global yield stagnation in cocoa production, the third largest trade commodity globally. An ideal solution would be climate-adapted and biodiversity-friendly cocoa agroforestry systems that can produce the high yields of monocultures in this pollination-limited crop. We use data of more than 150,000 farmers in Indonesia, Ghana, and Ivory Coast to show that manual pollination can help to produce high yields in agroforestry systems resilient to climate impacts. We propose a scenario, whereby manual pollination makes it possible to mitigate global cocoa production deficits, avoid negative long-term price effects, and to improve farmers' income and environmental sustainability.

**Introduction**

Demand for cocoa, the third largest trade commodity globally (*1*), is increasing in China, Russia, India, and Brazil by 2.5% per year (*2*), while production has declined by an average 1.5% annually for the past decade (*3*). Climate change is further exacerbating this trend through increasing drought events and pest outbreaks diminishing yields (*4*). The three major cocoa producing countries are Ivory Coast, Ghana, and Indonesia. In West Africa, 89.5% of the current cocoa production areas are predicted to experience a decline in cocoa suitability until 2050 (*5*, *6*). The biggest cocoa producing island Sulawesi in Indonesia will need strong climate adaptation strategies by 2050 (*7*). As yield declines also reduce the income of the 5 million cocoa farmers globally, small-scale producers either expand production into new, often forested areas or shift to alternative income sources to maintain their income (*8*). Stakeholders of the cocoa bean supply chain often advocate high-yielding monocultures to buffer climate-related yield declines and to avoid negative socioeconomic consequences for small-scale producers.

In cocoa monocultures, shade trees have been removed and chemical inputs (i.e., pesticide and fertilizer) are used to maintain high yields and profits in the short-term (*9*). Monocultures, however, negatively affect biodiversity and critical ecosystem services such as soil fertility, biological pest control, and pollination (*8*), and overall climate resilience (*8*, *10*). By contrast, agroforestry systems maintain or restore a shade tree cover above 40% (*11*), create a high climate resilience, require less chemical inputs, maintain essential ecosystem services (*12*, *13, 14*) and produce stable but lower yields than in monocultures (*9*, *10, 11*, *15*). Hence, the ideal system would be a Smart Agroforestry System (SAS), combining the high yields, so far known only from monocultures, with the climate-adaptation potential and high biodiversity value of agroforestry systems.

Companies and governments are advocating sustainable cocoa initiatives to avoid deforestation and to promote ecosystem services, biodiversity and sustainable livelihoods (*16*, *17*). For instance, the 2018 Cocoa and Forest Initiative commits producers to halt deforestation in cocoa supply chains and to support sustainable livelihoods (*17*). In addition, consumers are increasingly supporting eco-certification schemes that promote sustainable production practices (*18*, *19*). So far, these initiatives have failed to reconcile SAS with large-scale cocoa production.

Cocoa pollination may have the potential to overcome current yield deficits in climate resilient and sustainable production systems (*4*). Yield gaps due to pollination limitations have not been quantified (*20*) for the major producer countries, however, data on mean percentage pollination-to-fruit set ratio show that only 5-10% of the flowers develop into a fruit (and much less to a mature fruit (*20*)). This percentage is strongly limited by pollen deposition onto the flowers and hence, by effective pollination (*21*). Manual pollination is much more efficient to increase fruit set. In Indonesia, manual pollination experiments on all flowers on a tree in smallholder farms raised yields by 2.6 times compared to natural pollination, irrespective of fertilizer and pesticide inputs (*22*) (see also Supplementary Material SM1). The other two other experimental study relating manual pollination to cocoa yield, also from Indonesia, reported a 4-fold cocoa yield increase at only 40% manual pollination increase (*23*), and a 420% harvested fruits increase at 100% manual pollination per tree (*24*).

Given the pollination limitation of cocoa yield production and the urgency to identify SAS systems, an important question is how a large-scale adoption of manual pollination would affect the cocoa sector at the national and global levels.

Here, we quantify the effect of manual pollination in Ivory Coast, Ghana and Indonesia, which represent 66.8% of global cocoa bean production (*3*). We use detailed primary production and socioeconomic data (*22*, *25*, *26*) from all three countries in 2016, and long-term supply and demand elasticity of the global cocoa market to estimate the equilibrium impact of the supply shift from manual pollination as well as resulting environmental impacts. We recommend a solution based on manual pollination to maximize environmental and socio-economic benefits (Fig. 1).

**Materials and Methods**

*Household-level Data – West Africa & Indonesia*

For Ghana and Ivory Coast, we used the KIT household survey dataset, which comprises interview data of 3500 households in both countries for the 2016/2017 cocoa growing season. We use the available data on income structures and production costs. A detailed overview of the extensive sampling methods and dataset is provided at the KIT website (25). An overview of all relevant variables is provided in Tab. S1 and Tab. S2.

Our household data for Indonesia are based on 155,000 farmers in the Sustainable Cocoa Production Program (26) and of 28 farmers described in Toledo-Hernández et al. (22), where pollination experiments were conducted. Like for West Africa, we used the data on income structures and production costs obtained from social surveys. We also extracted means and an overview of all relevant variables is provided in Tab. S3.

*Hand pollination of cocoa flowers*

The hand pollination of cocoa flowers in farms of Central Sulawesi involves fives steps (Fig. S4): **Flower monitoring.** In the early morning, we counted all open flowers in each selected tree (hereafter "pollen-receptor tree") (Fig. S4A). Then, we calculated the number of flowers to be pollinated according to the assigned pollination rate. For example, if flower counts of a given tree was 100, and hand pollination rate assigned was 10%, then the number of flowers to be pollinated was 10. Then, we randomly picked and marked with pins and labels (with the pollination day and date) the flowers to be pollinated (Fig. S4B). **Hand pollination.** We hand pollinated cocoa flowers for 60 consecutive days following the methods described by Falque et al. (*41*) and Groeneveld et al. (*23*). First, we collected open flowers from three new trees (hereafter "pollen-source trees") in an additional farm not included in the study (Fig. S4C). This approach increases the genetic pool and avoids fruit abortion due to tree self-incompatibility. We randomly hand-picked one flower per pollen-source tree and carefully rubbed their anthers in the marked flower styles of the pollen-receptor tree (Fig. S4D). Finally, we removed flowers not pollinated to prevent open pollination. **Fruit set.** We recorded successful pollination, or fruit set, two days after hand pollination (Fig. S4E). This is because only successfully pollinated flowers remain on the tree 48h after pollination, while the unsuccessfully pollinated fall down (*42*). For example, we recorded fruit set of hand pollination in days one and 60 in days three and 62, respectively. **Fruit losses.** A large proportion of young fruits rotten and shrank in the first months of development. This phenomenon also known as cherelle wilt, or fruit abortion, is a plant regulating process associated with the limited plant energy resources available for fruit development (Fig. S4F), which eventually causes an early fruit abortion. We daily quantified fruit abortion for two weeks and later weekly until the harvest. We also weekly quantified fruit losses caused by the cocoa mosquito (*Helopeltis* sp.) pest and black pod disease (*Phytophthora* sp.) (Fig. S4G) until the harvest. **Harvest.** The harvest took place around six months after the hand pollination started. Here, we collected all harvestable fruits and quantified the proportion of healthy and diseased fruits. We opened the harvested fruits to extract the fresh beans and weighted them (fresh bean weight kg/tree) (Fig. S4H). We fermented and dried cocoa beans following the local practices consisting in a seven-days fermentation in rice sacks, and a two to three days open-sun drying (Fig. S4I). We quantified final yields as dry been weight (kg/tree) (Fig. S4J). More details are discussed in (22).

*Pollination Scenario Development and Cocoa Yield Changes*

In the 'no manual pollination' scenario, only the country-wide household level data above was considered to translate dry yields into farmer's income. In the 'intermediate manual pollination' and 'maximum manual pollination' scenarios, we use the pollination-yield multiplicator (PYM) factor of the most comprehensive pollination study from Indonesia ($PYM_{Min}$ = 2.6x; 212 cocoa trees on 28 farms) (22) and the average of all available pollination-yield effect studies ($PYM_{Max}$ = 3.3x; 260 cocoa trees on 34 farms)(*3*, *4*), respectively, to calculate income benefits for farmers in the three most important producer countries.

We calculate dry yield (Y) for the 'no manual pollination' ($Y_{No}$), 'intermediate manual pollination' ($Y_{Min}$), and 'maximum manual pollination' ($Y_{Max}$) scenarios:

***No manual pollination:*** $Y_{No} = Y$
***Intermediate manual pollination:*** $Y_{Min} = Y_{Dry} * PYM_{Min}$
***Maximum manual pollination:*** $Y_{Max} = Y_{Dry} * PYM_{Max}$

*Calculating long-term global cocoa price, supply, and farmer job changes*

As global cocoa production increases considering demand and supply elasticities, cocoa prices will decrease, which will affect demand. Some producers may leave the market because they cannot produce profitably. The new equilibrium is then at a lower price $\gamma_P$ and higher supply level $\gamma_S$, but it is weakened relative to the inelastic supply.

For subsequent price change calculations, we assume that all cocoa farms have a proportional yield increase of $\delta\%$ (we correct for realistic proportional pollination enhancement calculations below). Further the new price $P_N$ and the old price $P_O$ are in a constant relationship to each other

$$\frac{P_N}{P_O} = (1 + \delta\%)^{\frac{1}{\epsilon_D - \epsilon_S}} = \gamma_P$$

where $\epsilon_D$ is the demand elasticity $\epsilon_S$ denotes supply elasticity. $\gamma_P$ accounts for long-term demand and supply elasticity.

The old $S_O$ and new $S_N$ supply are also in a constant relationship defined as

$$\frac{S_N}{S_O} = (1 + \delta\%)^{\frac{\epsilon_D}{\epsilon_D - \epsilon_S}} = \gamma_S$$

The total supply change $\gamma_S$ at the global market is lower than the actual supply increase:

$$\lambda\% = \delta\% - (\gamma_S - 1)$$

$\lambda\%$ is the percentage of farmers that cannot produce crops at the reduced price and will find alternative income sources.

*Calculating short-term and long-term farmers income changes*

For all scenarios we used two cocoa prices. For short-term benefits without an immediate price response, $P_O = 2.28$ USD/kg, which is the global average cocoa price between 2001 and 2017. The assumption that all farms globally will adopt cocoa pollination enhancement methods is not realistic and, hence, we are comparing no manual pollination adoption (0%) to a realistic 25% adoption rate. We also present global production, supply, price and farmers job change estimations across adoption rates ranging from 0-100% in the Supplementary Information (Fig. S3). $P_N$ for the intermediate and maximum manual pollination scenario is $P_{N\_Min} = 1.89$ USD/kg and $P_{N\_Max} = 1.61$ USD/kg.

We calculate Gross Income ($I_{Gross}$) as:

***No manual pollination:*** $I_{Gross\_No} = Y_{No} * P_O$
***Intermediate manual pollination:*** $I_{Gross\_Min} = Y_{Min} * P_{N\_Min}$
***Maximum manual pollination:*** $I_{Gross\_Max} = Y_{Max} * P_{N_{Max}}$

The total operational costs ($OC_{Tot}$) are:

***No manual pollination:*** $OC_{Tot\_No} = Imp_{Tot} + Lab_{Farm}$
***Intermediate manual pollination:*** $OC_{Tot\_Min} = OC_{Poll} + Imp_{Tot} + Lab_{Farm}$
***Maximum manual pollination:*** $OC_{Tot\_Max} = OC_{Poll} + Imp_{Tot} + Lab_{Farm}$

Here Imp$_{Tot}$ are the total farm inputs including fertilizer, insecticide, herbicide, and fungicide costs. Lab$_{Farm}$ are the costs for farm based labour and also reflect the opportunity costs of the farmer. OC$_{Poll}$ is the labour cost to perform manual pollination, taking workers salary, the trees pollinated per day, and the days worked into account.

The net income (I$_{Net}$) is then:

$$\textbf{\textit{No manual pollination:}} \text{ I}_{Net\_No} = \text{I}_{Gross\_No} - \text{OC}_{Tot\_No}$$
$$\textbf{\textit{Intermediate manual pollination:}} \text{ I}_{Net\_Min} = \text{I}_{Gross\_Min} - \text{OC}_{Tot\_Min}$$
$$\textbf{\textit{Maximum manual pollination:}} \text{ I}_{Net\_Max} = \text{I}_{Gross\_Max} - \text{OC}_{Tot\_Max}$$

Note that we do not take changes in land value into account. See also Tab S1-S3.

**Results**

*Manual pollination effects on national and global cocoa production*

In the three major cocoa producing countries Ivory Coast, Ghana, and Indonesia, we find that 2016 farm level yields (averaged over all production systems) of respectively 273.2, 317.3, and 431.3 kg/ha with no manual pollination (hereafter 'No manual pollination' scenario) increase to 710.3, 824.9, and 1121,4 kg/ha with a 2.6 times manual pollination yield increase per tree (hereafter 'Intermediate manual pollination' scenario). Farm-level yields in the three countries can increase to 901.5, 1046.9, and 1423,3 kg/ha with a 330% manual pollination yield increase per tree (hereafter 'Maximum manual pollination' scenario; Tab. S1-S3). Hence, if manual pollination is adopted it will allow cocoa production with higher yields already at high shade levels, making it possible to design high-yielding and climate adapted agroforestry systems (for an example from Indonesia, see Fig. S1). If manual pollination had been adopted by 25% of all farmers in Ivory Coast, Ghana and Indonesia, the global production in the 2016/17 growing season would have increased by 18.3% and 38.1% in the intermediate and maximum manual pollination scenarios, respectively (Tab. 1). Under both scenarios, manual pollination has the potential to buffer global yield supply deficits and the predicted production declines of up to 1.2% due to pests and diseases (*27*).

*Manual pollination effects on the global cocoa market*

Short-term (i.e., weeks to months) supply and price effects of the adoption of manual pollination are negligible, because this practice is only adopted by a low percentage of farmers (Fig. S1; Fig. S2). Early-adopting farmers will benefit from the efficiency improvement minus the additional labour cost. In the long term however, and if global supply and demand elasticities of 0.57 and -0.34 respectively (*28*) remain constant, large-scale adoption of manual pollination would eventually lead to a larger supply and lower equilibrium price, produced by fewer and more efficient producers (Fig. S1; Fig. S2). The market equilibrium has annual global supply increases of 6.5% and 12.8% in the intermediate and maximum manual pollination scenarios, respectively (Tab. 1). Under these two scenarios, 11.8% and 25.3% of the original production would shift to alternative land uses and income sources (Tab. 1) and the equilibrium global cocoa price would decrease by 16.9% (1.89 USD/kg) and 29.9% (1.60 USD/kg), relative to the 2001-2017 mean cocoa price of 2.28 USD/kg (Tab. 1).

The resulting income changes depend on the time and labour costs required to perform manual pollination and on the price adjustment timeframe. If doubling the income per farmer is worth the effort without price adjustment (see Fig. 2 for an example of a 10% income increase), labour costs for manual pollination in Ivory Coast exceed these benefits after 30 and 80 pollination days per year in the minimum and maximum manual pollination scenarios, respectively. In Ghana and Indonesia for these two scenarios, manual pollination should not be done longer than 20 and 40 days, and 60 and 140 days in the respective countries. After price adjustment, these durations are reduced to 10 and 20 days in Ivory Coast and Indonesia and 10 days for both manual pollination scenarios in Ghana (Fig. 2). When comparing fixed labour expenses, say for 30 pollination days across all countries, the income gain is comparable between Indonesia and Ivory Coast, and lower in Ghana (Tab. S4). After price adjustment, the income gains are highest in Indonesia, followed by Ivory Coast and Ghana (Tab. S4).

*An alternative win-win manual pollination scenario*

We can envision a win-win scenario whereby the resultant yield increase from manual pollination would compensate for: (i) the predicted decline in cocoa suitability in some regions, (ii) a general shift away from monocultures to agroforestry in response to eco-certification standards, carbon markets, etc., and (iii) strict enforcement of the zero deforestation pledges made by most cocoa traders and chocolate makers (thus requiring intensification rather than expansion into forests; Fig. 3). We first assume that cocoa yield decreases by 40% (literature estimates for Indonesia, Ivory Coast, and Ghana are between 20-38% (*11*, *28*)) when 100% of production areas are converted from monocultures into agroforestry systems; second, habitat suitability decreases by 0.4% annually (a 9% decrease is reported across the global production areas until 2050 (*30*)) due to climate change, pest and diseases, and; third, there is zero-encroachment into new habitats (see Fig. 3 for other scenarios). Then, across Ivory Coast, Ghana, and Indonesia, 1,27 million tonnes of cocoa production can be compensated by manual cocoa pollination. With no increase in global production, the global cocoa price is not expected to decrease. Under such a win-win scenario, the sustainability of cocoa production and its resilience to climate shocks would increase; and cocoa production systems would deliver more ecosystem services (*8*, *12*).

**Discussion**

*The benefits and challenges of manual cocoa pollination*

We show how manual pollination can overcome pollination limitation of cocoa production and analyse the effects of this agricultural practice on global production, farmers' income, and cocoa price (Fig. 1). In the three major producer countries Ivory Coast, Ghana, and Indonesia, manual pollination increases yields with positive effects on global production and farmers' income. The anticipated supply changes in these three countries alone can outweigh global supply deficits by at least two-fold and, hence, satisfy increasing global demands. Higher adoption of manual pollination would lead to global supply increase and price declines, resulting in decreased socioeconomic and landscape-scale level benefits (Fig. 1). An increasing percentage of farmers would not be able to produce competitively at low prices, forcing them into alternative jobs, with a risk of eroding on-farm and landscape-level conservation benefits due to unsustainable land use change (*8*). In Ghana for example, farmers seek alternative employment opportunities in open-pit gold mining, which is already more lucrative than farming cocoa (*31*, *32*). Our win-win scenario, however, mitigates these negative long-term effects, because manual pollination is only adopted to compensate for a loss of suitable production area due to climate change and for yield losses caused by diseases and conversion from monocultures to climate-resilient agroforestry systems.

When manual pollination is used in agroforestry systems with shade cover above 40%, a biodiversity-friendly, and climate-adapted agroforestry system can produce comparable yields to monocultures. Locally, such a highly profitable system could risk stimulating expansion of cocoa plantations in new areas, therefore causing deforestation (see (*33*)). Such a rebound effect – where intensification leads to more expansion due to higher profitability – could be avoided through effective land-use zoning and implementation of zero deforestation pledges by the cocoa industry (*34*). Globally, however, enhancing efficiency with manual pollination will reduce the overall amount of land dedicated to cocoa if demand elasticity stays low. Actually, with a low demand elasticity, efficiency gains are likely to cause a land sparing effect.

*The future of Cocoa Production*

Climate change, diseases, and low global market prices are threatening global cocoa production. Manual pollination is a highly promising and immediate way to implement a win-win scenario to protect production, livelihoods, the environment, and help the industry and governments to fulfil their zero deforestation pledges in a highly pollination-limited production environment. In addition, interventions such as land use zoning, payments for ecosystem services, and creation of off-farm jobs are required to compensate potential negative effects on the economy and the environment.

Research should focus now on a better understanding of manual pollination. As profits resulting from manual pollination depend largely on operational costs, it is crucial to understand and optimize the number of days required for manual pollination (*24*). Systematization of manual pollination through preparation of pollen, pollinator sticks, or farmer training may reduce operational costs even further (*35*). Moreover, manual pollination may not necessarily be equally efficient in all cocoa producing countries and farming systems. For example, low shade levels may lead to a stress disposition of cocoa trees (*8*) resulting in reduced manual pollination potential in low shade monocultures (*20*). Performance of different cocoa genotypes under hand pollination should be also taken into consideration (*24*). Social acceptance of new production techniques are also likely to vary. Finally, more empirical work is needed to understand the long-term effects of manual pollination and open pollination on production, and the social, economic, and environmental benefits (*36*, *37*).

On the long-term, land-use policies, new varieties, and enhanced natural pollination — in addition to manual pollination – will ensure SAS goal. Land-use policies should encourage future SAS development based on multi-stakeholder initiatives and existing sustainability efforts (*38*). These policies should find a context-specific balance between levels of i) environmental protection, ii) management of cocoa production landscapes; and iii) planning for restoration of degraded production landscapes. New gene-edited varieties are less susceptible to climate change and more resistant to major cocoa pests (*39*). Different cocoa varieties in commercial plantations also maintain yields, e.g., due to avoidance of incompatibility (*40*). Little studied natural cocoa pollination is a low cost (it is free for the farmer) and less effective (natural pollination rate is about 10% (*18*)) alternative to manual pollination. It likely requires SAS systems with more than 40% canopy cover and moist and dark leaf litter habitats (*20*). If producing countries promote manual pollination, climate-resilient and biodiversity-friendly agroforestry systems have the potential to ensure a supply of cocoa in the decades to come. Eco-certification standards, carbon markets, and zero deforestation pledges by cocoa traders and chocolate makers can accelerate the transition to a more sustainable cocoa production system.

**Acknowledgments**

We thank Christina Rina and her team for providing the Indonesian cocoa farmer survey dataset. We also thank Anna Laven and her team at KIT for compiling and making the Ivory Coast and Ghana cocoa farmer survey dataset publicly available. TCW was supported by a Westlake University startup grant.

**Figures and Tables**

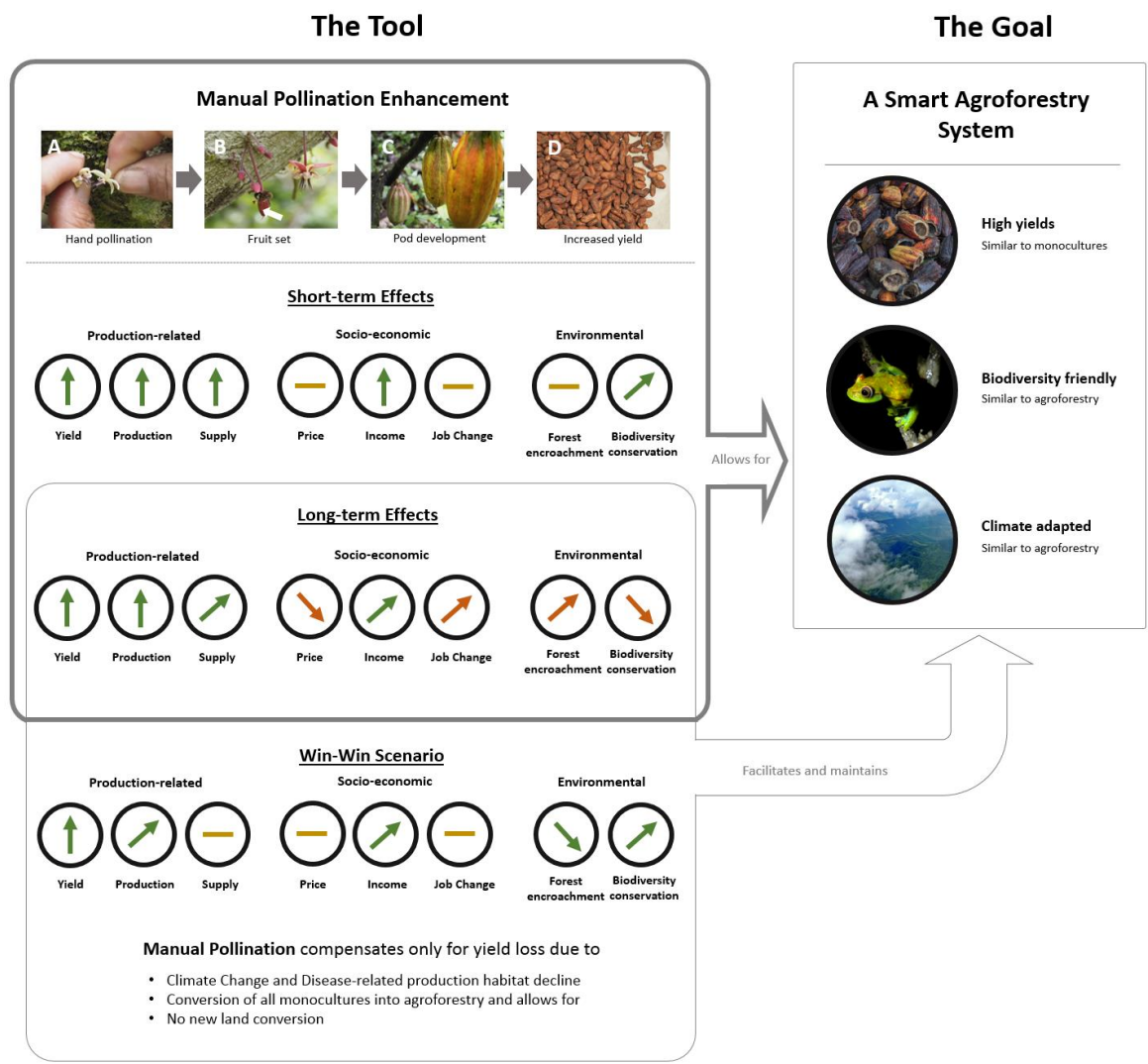

**Figure 1.** Manual pollination as a tool to prepare high yielding, biodiversity friendly, and climate adapted smart agroforestry systems (SAS). Manual pollination has both short- and long-term effects on production-related, socio-economic, and environmental effects (e.g., green upwards and red downward pointing arrows indicate a respective positive increase and negative decrease; a horizontal line indicates no change). The long-term negative effects on global cocoa price ('Price'), farmers' job choices ('Job change'), and environmental aspects can be mitigated with a win-win scenario, whereby only specific climate and disease-related yield losses and conversion of monocultures into agroforestry systems are compensated. This latter approach will help to achieve sustainable cocoa pledges by the chocolate industry and overall sustainable production systems.

## Short-term

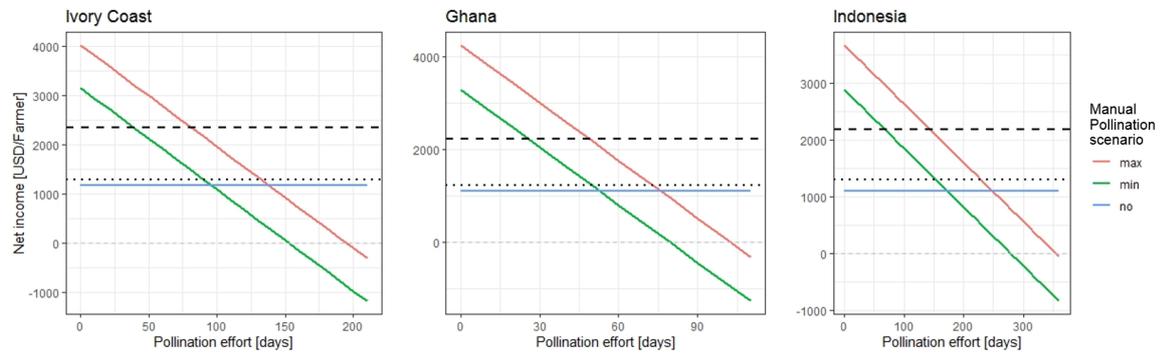

## Long-term

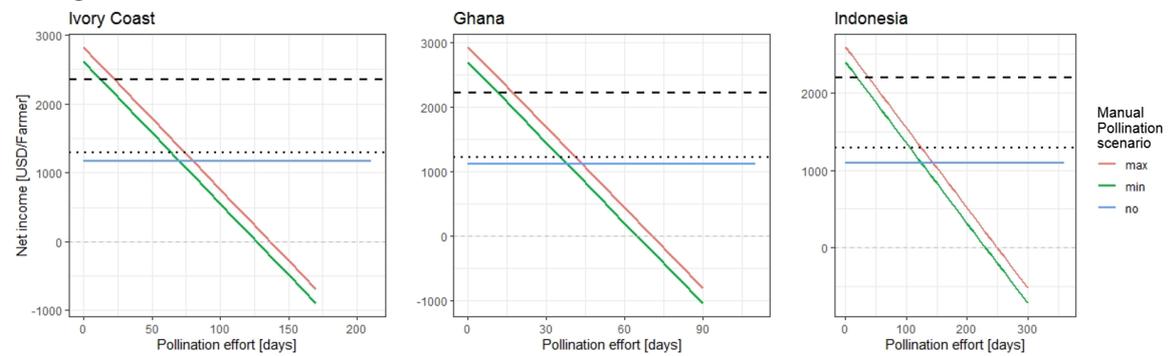

**Figure 2.** Short and long-term pollination effort (i.e., the amount of labour required for pollination activity) and the resulting net income. The black dashed and dotted lines represent the respective double and 10% income increase goals.

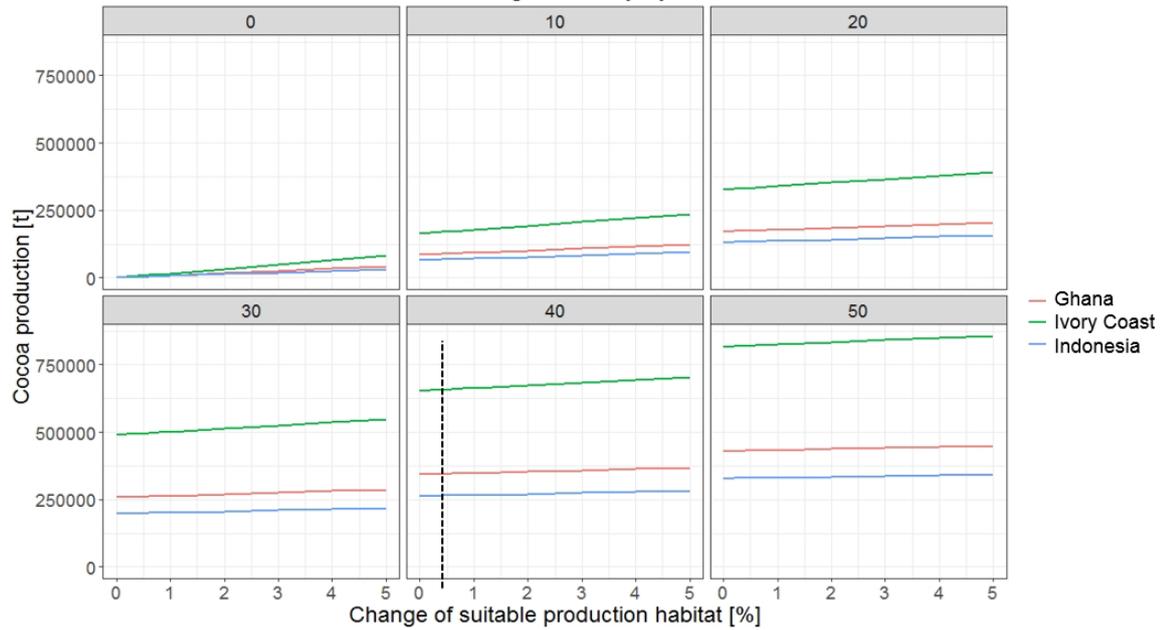

**Figure 3.** Quantifying the win-win scenario. Change of suitable production area and conversion of monocultures into agroforestry plantations. A realistic value for yield changes from monocultures to agroforestry is 40% and average annual change of suitable production is about 0.4% (dashed vertical line). The breadth of values illustrates potential variation of effects between countries.

**Table 1.** Production and socioeconomic effects of manual cocoa pollination.

|  | Minimum Scenario | Maximum Scenario |
|---|---|---|
| *25% pollination adoption rate* | 2.6x pollination increase | 3.3x pollination increase |
| **Global bean production [t]** | 4,466,574.00 ||
| Increased global bean production [t], (%) | 5,285,315.91 (18.3) | 6,168,740.32 (38.1) |
| Ivory Coast addition [t] | 311,555.06 | 759,415.45 |
| Ghana addition [t] | 213,669.78 | 520,820.09 |
| Indonesia addition [t] | 293,517.07 | 421,930.79 |
| Supply change [%] | 6.5 | 12.8 |
| Price change [USD], (%) | 1.89 (16.9) | 1.60 (29.9) |
| Farmer's job change [USD], [%] | 11.8 | 25.3 |

# Supplementary Information

Cocoa Pollination, biodiversity-friendly production, and the global market

Thomas Cherico Wanger[1,2,3,4*], Francis Dennig[5,6], Manuel Toledo-Hernández[1,2,3], Teja Tscharntke[2], Eric F. Lambin[7,8]

[1]Sustainability, Agriculture & Technology, School of Engineering, Westlake University, China. [2]Agroecology, University of Göttingen, Germany. [3]GlobalAgroforestryNetwork.org, China. [4]Key Laboratory of Coastal Environment and Resources of Zhejiang Province, Westlake University, China. [5]CASBS, Stanford University, USA. [6]Yale-NUS College, Singapore. [7]School of Earth, Energy & Environment Sciences and Woods Institute for the Environment, Stanford University, USA. [8]Georges Lemaître Earth and Climate Research Centre, Earth and Life Institute, Université catholique de Louvain, Belgium.

*Thomas Cherico Wanger

**Email:** tomcwanger@gmail.com

**This PDF file includes:**

Figs. S1 to S4

Tables S1 to S4

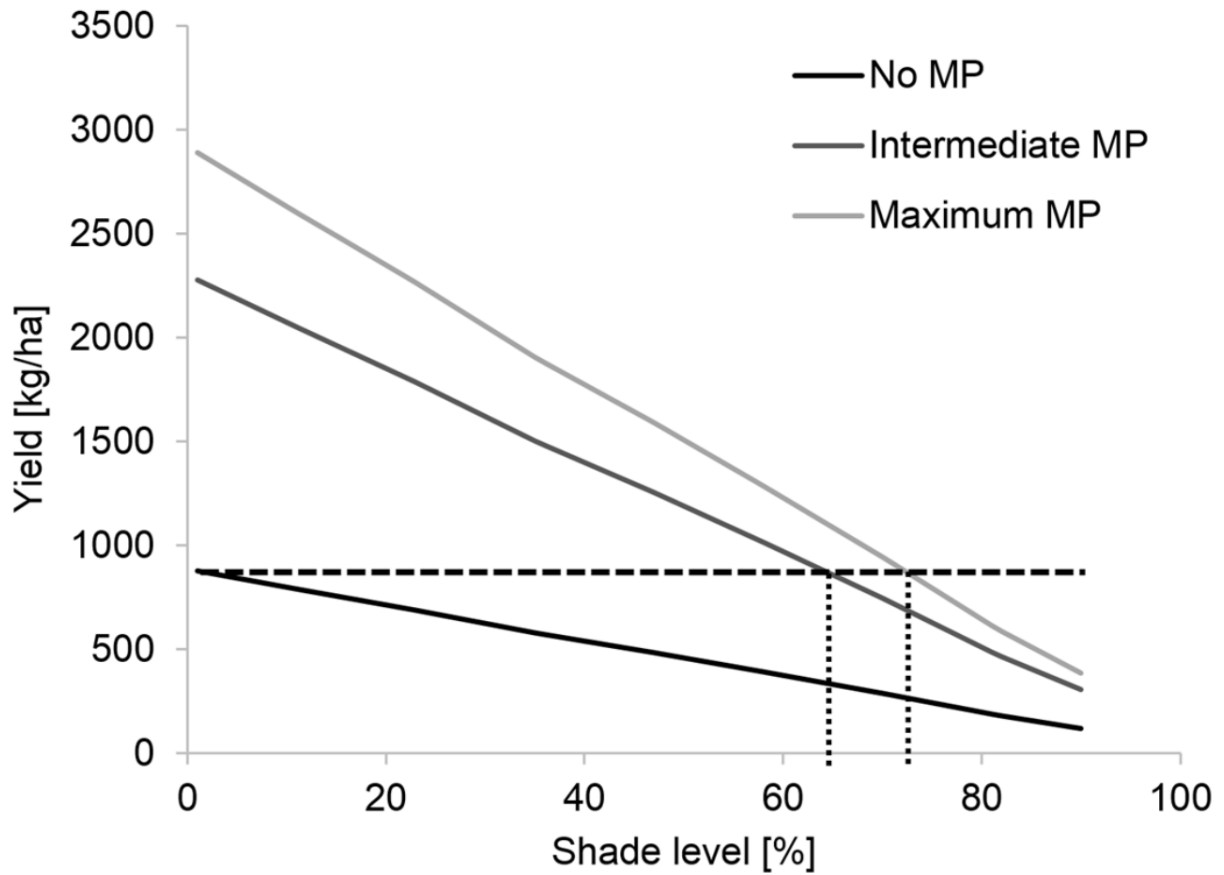

**Figure S1.**
Manual cocoa pollination (MP) effects on Indonesian cocoa production. Currently, the highest yields in monocultures are achieved at 0% shade cover (based on data in (*7*)). However, these high yields are paralleled by low climate adaptation potential and lead to biodiversity and ecosystem service decline (*8*). If MP is implemented, the maximum yields at 0% shade can already be achieved at respective 64% and 72% shade cover at intermediate and maximum MP scenarios (see Material and Methods for details).

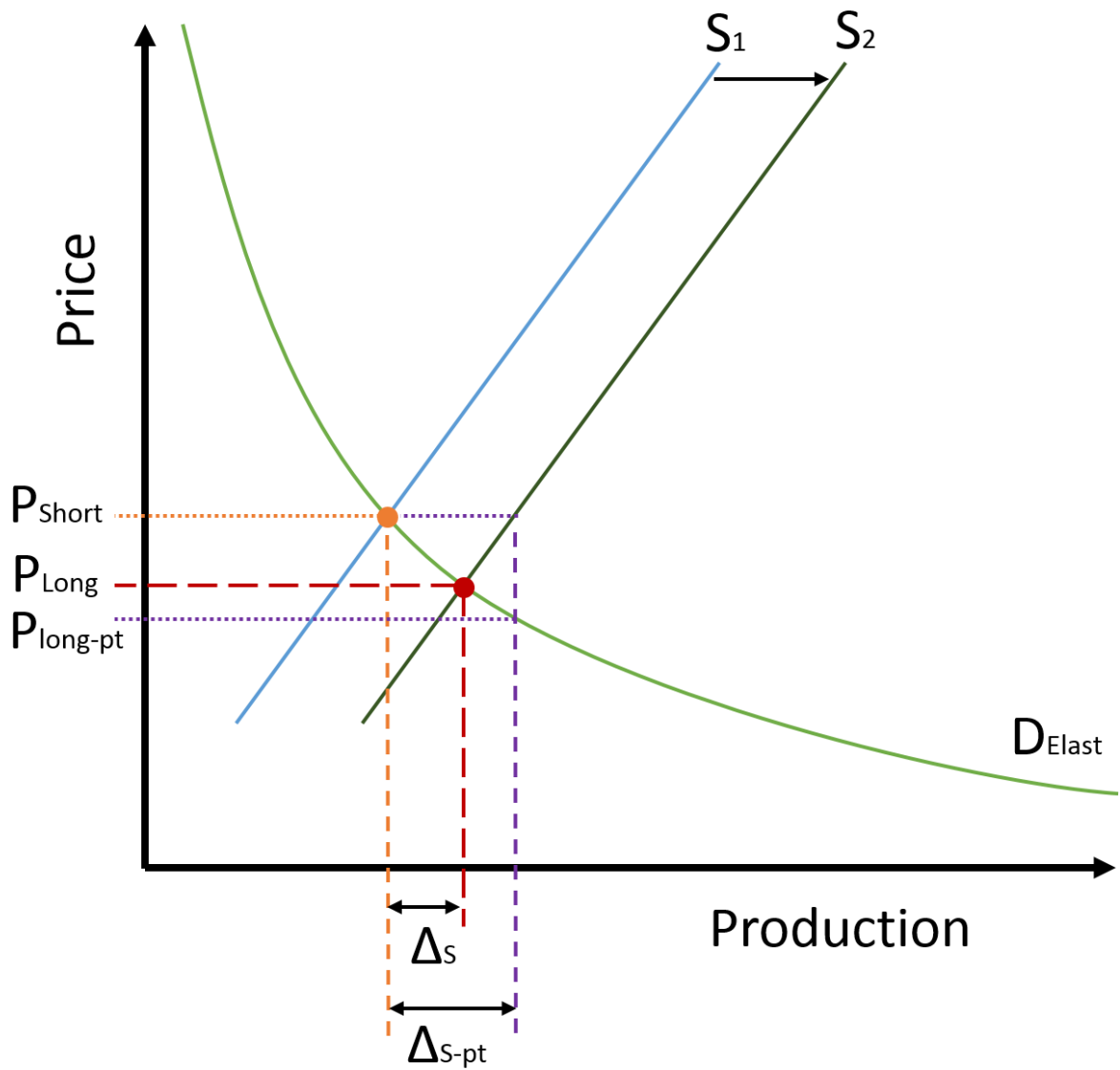

**Figure S2.**
Supply-Demand Curve considering low elasticity in the global cocoa market. In the short term, the equilibrium price (PShort) is fixed, because ∆S due to hand polination is small (orange point). In the long term, the global market equilibrium price (PLong) responds to hand pollination-related supply adjustments (shift from S1 to S2; red point). A low supply elasticity leads to a lower supply response (∆S-pt) when production is increased compared to a completely elastic price response (PLong-pt).

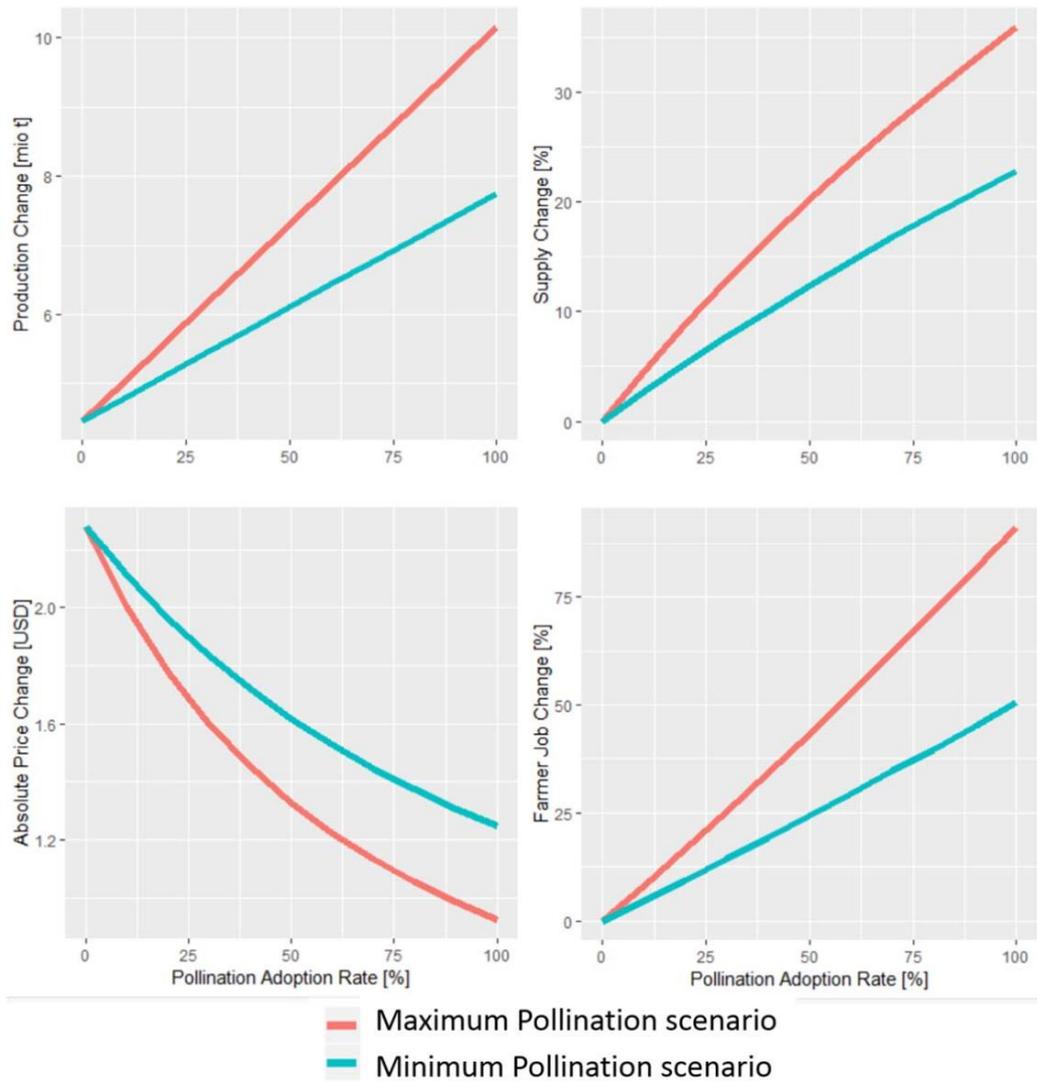

**Figure S3**. Global manual pollination effects for production and socioeconomic parameters across the entire pollination adoption range.

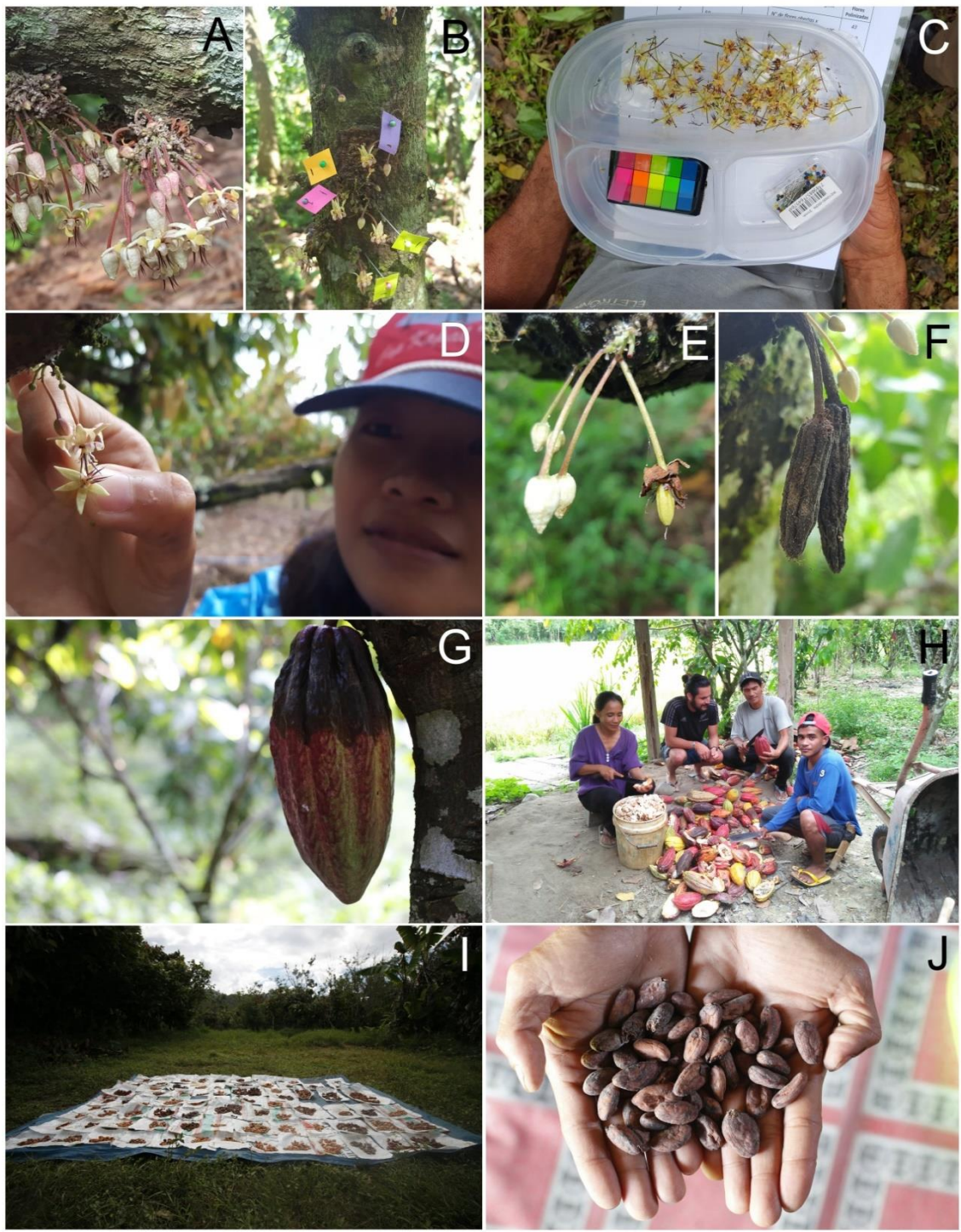

**Figure. S4.** The step by step hand pollination of cocoa flowers in farms of Central Sulawesi (see also Toledo-Hernández et al. in review). We quantified cocoa flowers in a selected tree (A) and marked the flowers for hand pollination (B). Then, we collected flowers from three trees in an additional farm (C), and used them to hand pollinate the marked flowers (D). After hand pollination, we quantified fruit set (E), cherelle wilt (F), and pest and diseases (G). We harvested mature fruits around six months after hand pollination started (H). For the fermentation and drying of beans we follow local practices (I). Finally, we recorded fresh and dry weigh (kg/tree) (J).

**Table S1.**

Ivory Coast – pollination effects on global cocoa production and farmers' income

| Variables | Parameter | Unit | 60 days | 30 days | Sources & Comments |
|---|---|---|---:|---:|---|
| Production | Total cocoa bean production_noPollination | tonnes | 4,466,574.00 | 4,466,574.00 | FAO |
| Production | Total cocoa bean production_minScenario | tonnes | 5,712,794.22 | 5,712,794.22 | |
| Production | Total cocoa bean production_maxScenario | tonnes | 7,504,235.79 | 7,504,235.79 | |
| Production | Yield_noPollination | tonnes | 778,887.64 | 778,887.64 | Percentage of global production produced by Ivory Coast |
| Production | Yield_minScenario | tonnes | 2,025,107.86 | 2,025,107.86 | |
| Production | Yield_maxScenario | tonnes | 2,570,329.21 | 2,570,329.21 | |
| Production | Minium scenario - pollination increase | | 2.60 | 2.60 | Toledo-Hernández et al. 2020 |
| Production | Maximum scenario - pollination increase | | 3.30 | 3.30 | Groeneveld et al. 2010 & Toledo-Hernández et al. 2020 |
| Production | Total area harvested 2016 | ha | 2,851,084.00 | 2,851,084.00 | FAO |
| Production | Number of Farmers | number | 1,000,000.00 | 1,000,000.00 | KIT Study |
| Production | Percentage produced by farmers | % | 0.70 | 0.70 | KIT Study |
| Production | trees_perHa | number | 975.00 | 975.00 | Daymond et al. 2017 |
| Gross income | YieldDry_maxScenario | kg/ha | 901.53 | 901.53 | |
| Gross income | YieldDry_minScenario | kg/ha | 710.29 | 710.29 | |
| Gross income | YieldDry_noPollination | kg/ha | 273.19 | 273.19 | KIT study // High season values used |
| Gross income | Mean Cocoa bean price 2001-2016 | USD/kg | 2.28 | 2.28 | ICCO data |
| **GrossIncome** | **Total Gross Income** | **USD/ha/year** | **622.87** | **622.87** | |
| **GrossIncome** | **Total Gross Income_poll** | **USD/ha/year** | **1,619.47** | **1,619.47** | |
| **GrossIncome** | **Total Gross Income_poll_mean** | **USD/ha/year** | **2,055.48** | **2,055.48** | |
| Operational cost | Fertilizer | USD/ha/year | 7.50 | 7.50 | KIT study |
| Operational cost | Insecticides | USD/ha/year | 10.72 | 10.72 | KIT study |
| Operational cost | Herbicide | USD/ha/year | 3.80 | 3.80 | KIT study |
| Operational cost | Fungicides | USD/ha/year | 0.60 | 0.60 | KIT study |
| Operational cost | Labour | USD/ha/year | 11.95 | 11.95 | KIT study |
| **OpCost_Farm** | **Total Farm OpCost** | **USD/ha/year** | **34.57** | **34.57** | |
| Operational cost | Salary per day | USD/10H | 0.82 | 0.82 | Mean of salaries per day |
| Operational cost | Trees per worker | number/day | 77.10 | 77.10 | Toledo-Hernández et al. 2020 |
| Operational cost | Days of work | days | 60.00 | 30.00 | Toledo-Hernández et al. 2020 |
| Operational cost | Total salary per hectar | USD/ha | 10.36 | 10.36 | Total number of trees per ha / trees per worker per day * salary per day |
| **OpCost_Pollination** | **Total Pollination OpCost** | **USD/ha** | **621.35** | **310.67** | |
| NetIncome | Total Net Income | USD/ha | 588.30 | 588.30 | |
| NetIncome | Total Net Income_minScenario | USD/ha | 963.55 | 1,274.23 | |
| NetIncome | Total Net Income_maxScenario | USD/ha | 1,399.56 | 1,710.24 | |
| NetIncome | National Income | USD | 1,174,111,288.47 | 1,174,111,288.47 | |
| NetIncome | National Income_minScenario | USD | 1,923,015,001.82 | 2,543,046,881.29 | |
| NetIncome | National income_maxScenario | USD | 2,793,188,270.95 | 3,413,220,150.42 | |
| NetIncome | Farmer income | USD / Farmer | 1,174.11 | 1,174.11 | |
| NetIncome | Farmer Income_minScenarion | USD / Farmer | 1,923.02 | 2,543.05 | |
| NetIncome | Farmer income_maxScenario | USD / Farmer | 2,793.19 | 3,413.22 | |

**Table S2.**

Ghana – pollination effects on global cocoa production and farmers' income

| Variables | Parameter | Unit | 60 days | 30 days | Sources & Comments |
|---|---|---|---|---|---|
| Production | Total cocoa bean production_noPollination | tonnes | 4,466,574.00 | 4,466,574.00 | FAO |
| Production | Total cocoa bean production_minScenario | tonnes | 5,321,253.11 | 5,321,253.11 | |
| Production | Total cocoa bean production_maxScenario | tonnes | 6,549,854.34 | 6,549,854.34 | |
| Production | Yield_noPollination | tonnes | 534,174.45 | 534,174.45 | Percentage of global production produced by Ghana |
| Production | Yield_minScenario | tonnes | 1,388,853.56 | 1,388,853.56 | |
| Production | Yield_maxScenario | tonnes | 1,762,775.67 | 1,762,775.67 | |
| Production | Minium scenario - pollination increase | NA | 2.60 | 2.60 | Toledo-Hernández et al. 2020 |
| Production | Maximum scenario - pollination increase | NA | 3.30 | 3.30 | Groeneveld et al. 2010 & Toledo-Hernández et al. 2020 |
| Production | Total area harvested 2016 | ha | 1,683,765.00 | 1,683,765.00 | FAO |
| Production | Number of Farmers | number | 800,000.00 | 800,000.00 | KIT Study |
| Production | Percentage produced by farmers | % | 0.90 | 0.90 | FAO |
| Production | trees_perHa | number | 1,244.00 | 1,244.00 | Daymond et al. 2017 |
| Gross income | YieldDry_maxScenario | kg/ha | 1,046.93 | 1,046.93 | |
| Gross income | YieldDry_minScenario | kg/ha | 824.85 | 824.85 | |
| Gross income | YieldDry_noPollination | kg/ha | 317.25 | 317.25 | KIT study // High season values used |
| Gross income | Mean Cocoa bean price 2001-2016 | USD/kg | 2.28 | 2.28 | ICCO data |
| **Gross income** | **Total Gross Income** | **USD/ha/year** | **723.33** | **723.33** | |
| **Gross income** | **Total Gross Income_minScenario** | **USD/ha/year** | **1,880.66** | **1,880.66** | |
| **Gross income** | **Total Gross Income_maxScenario** | **USD/ha/year** | **2,386.99** | **2,386.99** | |
| Operational cost | Fertilizer | USD/ha/year | 10.59 | 10.59 | KIT study |
| Operational cost | Insecticides | USD/ha/year | 18.79 | 18.79 | KIT study |
| Operational cost | Herbicide | USD/ha/year | 8.86 | 8.86 | KIT study |
| Operational cost | Fungicides | USD/ha/year | 2.65 | 2.65 | KIT study |
| Operational cost | Labour | USD/ha/year | 96.68 | 96.68 | KIT study |
| **OpCost_Farm** | **Total Farm OpCost** | **USD/ha/year** | **137.57** | **137.57** | |
| Operational cost | Salary per day | USD/10H | 1.36 | 1.36 | Mean of salaries per day |
| Operational cost | Trees per worker | number/day | 77.10 | 77.10 | Toledo-Hernández et al. 2020 |
| Operational cost | Days of work | days | 60.00 | 30.00 | Toledo-Hernández et al. 2020 |
| Operational cost | Total salary per hectar | USD/ha | 21.97 | 21.97 | Total number of trees per ha / trees per worker per day * salary per day |
| **OpCost_Pollination** | **Total Pollination OpCost** | **USD/ha** | **1,318.28** | **659.14** | |
| NetIncome | Total Net Income | USD/ha | 585.76 | 585.76 | |
| NetIncome | Total Net Income_minScenario | USD/ha | 424.81 | 1,083.95 | |
| NetIncome | Total Net Income_maxScenario | USD/ha | 931.14 | 1,590.28 | |
| NetIncome | National Income | USD | 887,653,967.76 | 887,653,967.76 | |
| NetIncome | National Income_minScenario | USD | 643,745,572.95 | 1,642,600,541.32 | |
| NetIncome | National income_maxScenario | USD | 1,411,033,747.54 | 2,409,888,715.91 | |
| NetIncome | Farmer income | USD / Farmer | 1,109.57 | 1,109.57 | |
| NetIncome | Farmer Income_minScenarion | USD / Farmer | 804.68 | 2,053.25 | |
| NetIncome | Farmer income_maxScenario | USD / Farmer | 1,763.79 | 3,012.36 | |

**Table S3.**

Indonesia – pollination effects on global cocoa production and farmers' income

| Variables | Parameter | Unit | 60 days | 30 days | Sources & Comments |
|---|---|---|---|---|---|
| Production | Total cocoa bean production_noPollination | tonnes | 4,466,574.00 | 4,466,574.00 | FAO |
| Production | Total cocoa bean production_minScenario | tonnes | 5,640,642.30 | 5,640,642.30 | |
| Production | Total cocoa bean production_maxScenario | tonnes | 6,154,297.18 | 6,154,297.18 | |
| Production | Yield_noPollination | tonnes | 733,792.69 | 733,792.69 | Percentage of global production produced by Indonesia |
| Production | Yield_minScenario | tonnes | 1,907,860.98 | 1,907,860.98 | |
| Production | Yield_maxScenario | tonnes | 2,421,515.86 | 2,421,515.86 | |
| Production | Minium scenario - pollination increase | | 2.60 | 2.60 | Toledo-Hernández et al. 2020 |
| Production | Maximum scenario - pollination increase | | 3.30 | 3.30 | Groeneveld et al. 2010 & Toledo-Hernández et al. 2020 |
| Production | Total area harvested 2016 | ha | 1,701,351.00 | 1,701,351.00 | FAOSTATS |
| Production | Percentage produced by farmers | % | 0.94 | 0.94 | SwissContact |
| Production | trees_ha | number | 738.25 | 738.25 | ISCR 2017 |
| Gross income | YieldDry_maxScenario | kg/ha | 1,423.29 | 1,423.29 | |
| Gross income | YieldDry_minScenario | kg/ha | 1,121.38 | 1,121.38 | |
| Gross income | YieldDry_noPollination | kg/ha | 431.30 | 431.30 | SwissContact |
| Gross income | Mean Cocoa bean price 2001-2016 | USD/kg | 2.28 | 2.28 | ICCO 2018 |
| **Gross income** | **Total Gross Income** | **USD/ha/year** | **983.36** | **983.36** | |
| **Gross income** | **Total Gross Income_minScenario** | **USD/ha/year** | **2,556.75** | **2,556.75** | |
| **Gross income** | **Total Gross Income_maxScenario** | **USD/ha/year** | **3,245.10** | **3,245.10** | |
| Operational cost | Fertilizer | USD/ha/year | 6.18 | 6.18 | SwissContact |
| Operational cost | Insecticide | USD/ha/year | 6.13 | 6.13 | SwissContact |
| Operational cost | Fungicide | USD/ha/year | 2.02 | 2.02 | SwissContact |
| Operational cost | Herbicide | USD/ha/year | 5.65 | 5.65 | SwissContact |
| Operational cost | Labor | USD/ha/year | 4.36 | 4.36 | SwissContact |
| **OpCost_Farm** | **Total Farm OpCost** | **USD/ha/year** | **24.34** | **24.34** | |
| Operational cost | Salary per day | USD/9H | 0.95 | 0.95 | SwissContact |
| Operational cost | Tree per worker | number/days | 77.10 | 77.10 | Toledo-Hernández et al. 2020 |
| Operational cost | Total salary per ha | USD/ha | 9.10 | 9.10 | SwissContact |
| Operational cost | Days of work | days | 60.00 | 30.00 | Toledo-Hernández et al. 2020 |
| **OpCost_Pollination** | **Total Pollination OpCost** | **USD/ha** | **545.79** | **272.89** | |
| NetIncome | Total Net Income | USD/ha | 959.02 | 959.02 | |
| NetIncome | Total Net Income_minScenario | USD/ha | 1,986.62 | 2,259.51 | |
| NetIncome | Total Net Income_maxScenario | USD/ha | 2,674.97 | 2,947.87 | |
| NetIncome | National Income | USD | 1,533,738,254.94 | 1,533,738,254.94 | |
| NetIncome | National Income_minScenario | USD | 3,177,139,188.94 | 3,613,570,310.16 | |
| NetIncome | National income_maxScenario | USD | 4,278,004,328.63 | 4,714,435,449.85 | |
| NetIncome | Farmer income | USD | 1,095.53 | 1,095.53 | |
| NetIncome | Farmer Income_minScenarion | USD | 2,269.39 | 2,581.12 | |
| NetIncome | Farmer income_maxScenario | USD | 3,055.72 | 3,367.45 | |

**Table S4.**

Short and long-term income effects of no, intermediate, and maximum manual pollination changes. Comparisons are based on the 2001-2017 mean price of 2.28 USD/kg; the top panel assumes no immediate market adjustment to increased supplies, the lower panel uses 1.89 USD/kg and 1.61 USD/kg for the intermediate and maximum manual pollination scenarios, respectively. Both panels assume a 25% pollination adoption rate. Dark green cells show a doubled income, light green cells indicate an increased income, and orange cells show a decreasing income. Values in brackets indicate percentage changes.

**Short-term**

| Pollination Scenario | Pollination Duration [days] | Ivory Coast [USD/ha] | Ghana [USD/ha] | Indonesia [USD/ha] |
|---|---|---|---|---|
| No Pollination | NA | 1,174.11 | 1,109.57 | 1,095.53 |
| Minimum | 30 | 2543.05 (116.6) | 2053.25 (85.0) | 2581.12 (135.6) |
| Minimum | 60 | 1923.02 (63.8) | 804.68 (-27.5) | 2269.39 (107.2) |
| Maximum | 30 | 3413.22 (190.7) | 3012.36 (171.5) | 3367.45 (207.4) |
| Maximum | 60 | 2793.19 (137.9) | 1763.79 (59.0) | 3055.72 (178.9) |

**Long-term**

| Pollination Scenario | Pollination Duration [days] | Ivory Coast [USD/ha] | Ghana [USD/ha] | Indonesia [USD/ha] |
|---|---|---|---|---|
| No Pollination | NA | 1,174.11 | 1,109.57 | 1,095.53 |
| Minimum | 30 | 1990.19 (69.5) | 1443.89 (30.1) | 2081.53 (90.0) |
| Minimum | 60 | 1370.16 (16.7) | 195.32 (-82.4) | 1769.80 (61.5) |
| Maximum | 30 | 2207.74 (88.0) | 1683.67 (51.7) | 2733.36 (149.5) |
| Maximum | 60 | 1587.70 (35.2) | 435.10 (-60.8) | 1966.38 (79.5) |